\documentclass[12pt]{iopart}

\usepackage{graphicx}
\usepackage{iopams} 
\usepackage{setstack} 
\usepackage{color}
\usepackage{multirow}

\usepackage[hidelinks]{hyperref}
\usepackage[all]{hypcap} 

\begin{document}

\title[]{Measurement of ($\alpha$,n) reaction cross sections of erbium isotopes for testing astrophysical rate
predictions}

\author{G.\,G.\,Kiss$^{1}$\footnote{Corresponding 
author: ggkiss@atomki.mta.hu}, T.\,Sz\"ucs$^{1}$\footnote{present address: Helmholtz-Zentrum Dresden-Rossendorf (HZDR), 01328 Dresden, Germany}, T.\,Rauscher$^{2,3}$, Zs.\,T\"or\"ok$^{1}$, L.\,Csedreki$^{1}$, Zs.\,F\"ul\"op$^{1}$, Gy.\,Gy\"urky$^{1}$, Z.\,Hal\'asz$^{1,4}$}

\address{$^1$ MTA Atomki, H-4001 Debrecen, POB. 51, Hungary}
\address{$^2$ Centre for Astrophysics Research, School of Physics, Astronomy and Mathematics, University of Hertfordshire, Hatfield AL10 9AB, United Kingdom}
\address{$^3$ Department of Physics, University of Basel, CH-4056 Basel, Switzerland}
\address{$^4$University of Debrecen, Department of Theoretical Physics, H-4001 Debrecen, Hungary }

\begin{abstract}
The $\gamma$-process in core-collapse and/or type Ia supernova explosions is thought to explain the origin of the majority of the so-called $p$ nuclei (the 35 proton-rich isotopes between Se and Hg). Reaction rates for $\gamma$-process reaction network studies have to be predicted using Hauser-Feshbach statistical model calculations. Recent investigations have shown problems in the prediction of $\alpha$-widths at astrophysical energies which are an essential input for the statistical model. It has an impact on the reliability of abundance predictions in the upper mass range of the $p$ nuclei. With the measurement of the $^{164,166}$Er($\alpha$,n)$^{167,169}$Yb reaction cross sections at energies close to the astrophysically relevant energy range we tested the recently suggested low energy modification of the $\alpha$+nucleus optical potential in a mass region where $\gamma$-process calculations exhibit an underproduction of the $p$ nuclei.
Using the same optical potential for the $\alpha$-width which was derived from combined $^{162}$Er($\alpha$,n) and $^{162}$Er($\alpha$,$\gamma$) measurement makes it plausible that a low-energy modification of the optical $\alpha$+nucleus potential is needed.

\end{abstract}

\vspace{2pc}

\submitto{\jpg}

\maketitle

\section{Introduction}
\label{sec:int}

Contrary to the synthesis of the light nuclei -- where charged particles are playing a particularly important role \cite{ilibook} -- two different neutron-capture processes, the so-called $s$ \cite{kap11} and $r$ \cite{arcreview} processes, are required to produce the bulk of naturally occurring nuclides above Fe. These two processes were found to be unable, however, to create 35 neutron-deficient, stable, rare isotopes between $^{74}$Se and $^{196}$Hg, which were called ``$p$ nuclei'' or ``excluded isotopes'' \cite{b2fh,cam57}. Photodisintegration of stable nuclei either in the outer shells of massive stars during a core-collapse supernova explosion \cite{woo78,arn03} or during the explosion of a White Dwarf \cite{tra11} has been suggested as a production mechanism for most of these nuclei \footnote{The origin of few isotopes --- such as the $^{164}$Er,$^{152}$Gd and $^{180}$Ta nuclei --- is still under debate. Recently it was shown that there is a large $s$ process contribution to their observed abundances \cite{arl99, rau_rew}.}. Such a so-called $\gamma$-process commences at 2-3\,GK temperatures by sequences of ($\gamma$,n) reactions, which are replaced with ($\gamma$,p) and ($\gamma,\alpha$) reactions when reaching sufficiently neutron deficient nuclides in an isotopic chain. The whole process lasts only a few seconds, before the environment becomes too cool for the photodisintegrations. Recent calculations, however, are under-predicting the heavy $p$-nucleus abundances in the mass region between $150\leq A\leq 165$ \cite{rau02, rau_rew, ray95}. It remains unclear whether this deficiency is due to nuclear cross sections, stellar physics, or if alternative or additional processes have to be invoked \cite{rau_rew}.

In order to resolve this ambiguity, on one hand, improvements on the description of the astrophysical conditions under which the process takes place (seed isotope abundances, peak temperatures, time scale, etc.) are needed. On the other hand, uncertainties are introduced into the calculations by nuclear physics input, in particular by the reaction rates. The $\gamma$-process models require the use of huge reaction networks including ten thousands of nuclear reactions and the rates of these reactions at a given stellar temperature have to be known. The reaction rates are generally taken from calculations using the Hauser-Feshbach (H-F) statistical model \cite{hau52,rau97,rau00, rau01,rau11}. The accuracy of the H-F predictions mainly depends on the adopted nuclear models for the proton-, neutron-, $\alpha$-, and $\gamma$-widths. 
The predictions can be tested for only a limited number of reactions, since experimental data in the relevant mass and energy range are very scarce \cite{kadonis}. Although photodisintegrations play a role in the $\gamma$-process, their rates are computed from capture rates using detailed balance. Experimental information on quantities (such as particle- and $\gamma$-widths) for the calculation of the stellar rates (which cannot be directly measured, with a few exceptions\cite{rau13exc}) can be obtained from the study of capture reactions. 
This approach is not only technically less challenging, but also provides more relevant astrophysical information than the direct study of the $\gamma$ induced reactions \cite{rau11,moh07,rau12lett,rau_sensi,rau14}.

\begin{center}
\begin{figure*}
\resizebox{1\columnwidth}{!}{\rotatebox{0}{\includegraphics[clip=]{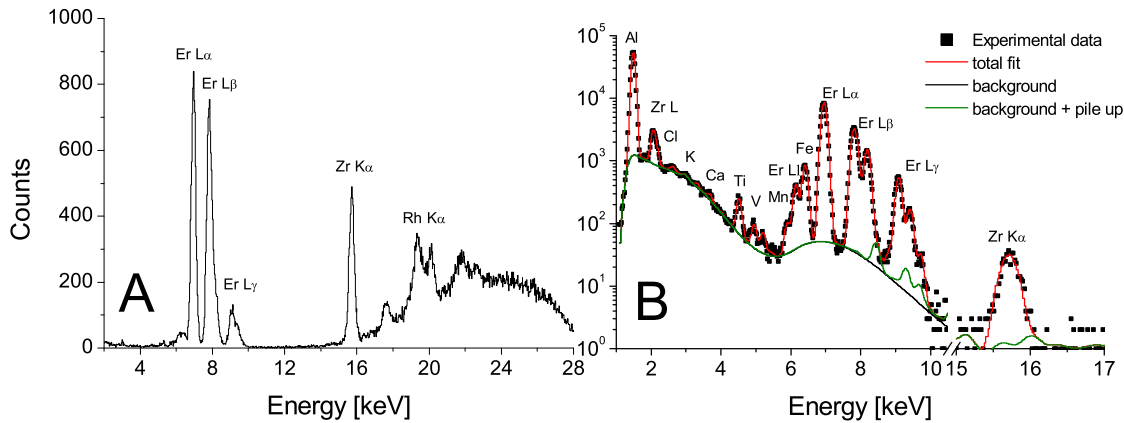}}}
\caption{\label{fig:target} (Color online) XRF (A) and PIXE (B) spectra used to determine the impurities and the thickness of the erbium targets. The peaks used for the analysis are marked. Peaks belonging to the Rh anode of the X-ray tube, to impurities in the target and/or the backing are indicated, too. For details, see text.}
\end{figure*}
\end{center}

Reactions with $\alpha$ particle absorbtion or emission in the $\gamma$-process are mainly relevant above the neutron shell closure at $N=82$ \cite{rau_rew,rau14}. Previous $\gamma$-process related experiments found discrepancies between measured and predicted $\alpha$-induced cross sections, which suggested low-energy modifications of the $\alpha$+nucleus optical potential \cite{kis11b,rau12,sau11,net13, glo14, kis14} or the need to modify the reaction model \cite{raulett}. An important recent break-through was the simultaneous, high-precision measurement of the $^{162}$Er($\alpha$,$\gamma$) and $^{162}$Er($\alpha$,n) reaction cross sections in a wide energy range below the Coulomb barrier \cite{kis14}. It was crucial for the theoretical interpretation that ($\alpha$,$\gamma$) data were taken even below the ($\alpha$,n) threshold. Thus, for the first time it was possible to show that the $\alpha$+nucleus optical potential requires an energy-dependent modification consistently and unambiguously within the same measurement at high masses. The aim of the present work is to provide further data in the same energy region for other Er isotopes and to test whether they can be described by the same approach as the $^{162}$Er data.

The paper is organized as follows: Experimental details are
described in Sec.~\ref{sec:exp}; the resulting cross sections are presented
in Sec.~\ref{sec:res}, where also a comparison with theoretical predictions is made;
Sec.~\ref{sec:sum} provides the conclusions and a short summary.

\section{Experimental technique}
\label{sec:exp}

The experiment was carried out at the Institute for Nuclear Research, Hungarian Academy of Sciences (MTA Atomki, Debrecen, Hungary). 
The element Er has six stable isotopes with mass numbers A\,=\,162, 164, 166, 167, 168 and 170. Alpha induced cross sections on the $p$ nucleus $^{162}$Er were already measured \cite{kis14}, the aim of the present work was to study the $^{164}$Er($\alpha$,n)$^{167}$Yb and $^{166}$Er($\alpha$,n)$^{169}$Yb reactions. To determine the $\alpha$-induced reaction cross sections the activation method was used; the $\gamma$-rays following the electron capture decay of the Yb reaction products were detected using a Low Energy Photon Spectrometer (LEPS) and a coaxial HPGe detector. The following paragraphs provide a detailed description of the experimental setup.

\begin{center}
\begin{figure*}                                                                                         
\resizebox{1.0\columnwidth}{!}{\rotatebox{0}{\includegraphics[clip=]{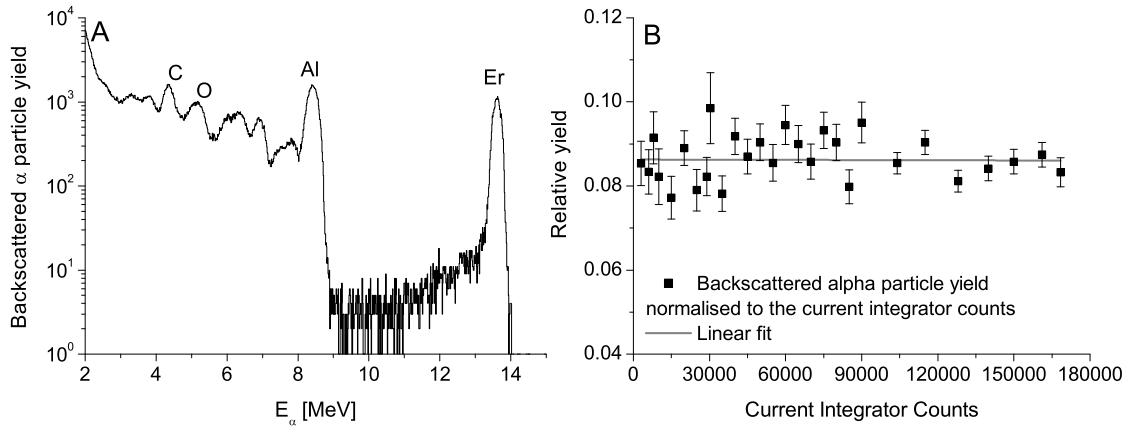}}}
\caption{\label{fig:rbs} Particle spectrum used to monitor the target thickness (A). The peaks corresponding to alpha backscattering on erbium, on the aluminum backing and on target impurities are indicated. On the right side the backscattering yield as the function of the current integrator counts is shown (B), a linear function was fitted to the data to prove that the target is stable. For more details see the text.}
\end{figure*}
\end{center}

\subsection{Target production, characterization}
\label{sec:tar}

The targets were made by vacuum evaporation of Er$_2$O$_3$ powder with natural isotopic composition ((1.61\,$\pm$\,0.03)\,\% $^{164}$Er and (33.6\,$\pm$\,3.5)\,\% $^{166}$Er, respectively) and enriched to (8.1\,$\pm$\,0.2)\% in $^{164}$Er ($^{166}$Er content: (33.10\,$\pm$\,3.4)\,\%) onto 2 $\mu$m thick, high purity Al foils. To check the enrichment provided by the supplier, at E$_{\alpha}$ = 15.5, 16.5 and 17.0\,MeV the irradiation was carried out using both natural and enriched targets and the derived cross sections were found to be consistent.

The Er targets were produced using reductive evaporation technique, the Er$_2$O$_3$ powder was mixed with Zr powder and placed into a carbon crucible heated by electron beam. During the evaporation, the distance between the Al backing and the evaporation boat was 8\,cm, using this distance, uniform targets could be produced. Care was taken to limit the target's Zn contamination --- identified as the main contaminant of the backing and the Zr powder used for the reduction --- since during the $\alpha$ irradiation $^{67}$Ge could be produced via the $^{64}$Zn($\alpha$,n)$^{67}$Ge reaction. This isotope (and its daughter) was identified as the main disturbing activity since during the $\beta$ decay a 208.95\,keV $\gamma$ ray is emitted which is only 1.15\,keV above the 207.80\,keV $\gamma$ of our interest (see Table \ref{tab:decay}). 
As a first step, the quality of the targets were tested by X-ray fluorescence spectroscopy (XRF) \cite{xrf}.
The targets were excited by filtered primary radiation (with a 2\,mm Al filter) of an Oxford Instruments X-ray tube (XTF 6000 BR) having Rh target. It was operated at 30\,kV voltage and 700\,$\mu$A electron beam current. The measurement time was 1000\,s in each case. The characteristic X-rays were detected in atmospheric pressure condition enabling the detection of trace elements down to potassium with high sensitivity. The evaluation of the spectra was carried out with the AXIL software \cite{axil}.

The absolute thickness of the targets were measured by weighing and using the PIXE (particle induced X-ray emission) technique \cite{kol11}. The PIXE chamber is installed on the left 45$^{\circ}$ beam line of the 5\,MV VdG accelerator of Atomki. A more detailed description of the PIXE setup can be found in Ref. \cite{ker10}. A homogeneous beam of 2\,MeV protons with a 5\,mm diameter and a 4-5\,nA current was used for the thickness measurement.
The total collected charge in the case of each target was about 2\,$\mu$C. The measured spectra were fitted using the PIXEKLM code \cite{szab06}.  
Typical XRF and PIXE spectra can be seen in Figure~\ref{fig:target} where peaks used for the analysis are marked.
From the analysis of the PIXE spectra the erbium target thickness was determined and found to be in
agreement with the weighing within maximum of 4.4\,\%. This 4.4\,\% --- as a conservative estimate --- has been adopted as the uncertainty of the target thickness determination.

\begin{center}
\begin{figure*}
\resizebox{1.0\columnwidth}{!}{\rotatebox{0}{\includegraphics[clip=]{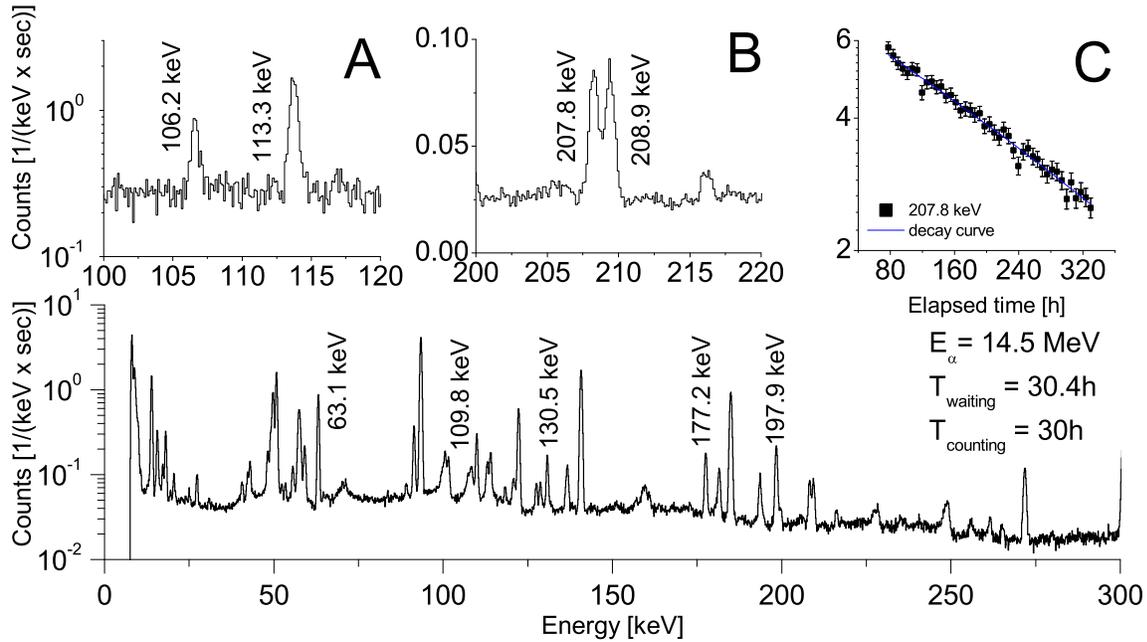}}}
\caption{\label{fig:spectra} Off-line $\gamma$ spectra, taken after the irradiation of an erbium target with 14.5\,MeV $\alpha$ beam.
The upper figures show the $\gamma$ lines used to determine the cross section of the $^{164}$Er($\alpha$,n)$^{167}$Yb reaction: in figure A the $\gamma$ transitions emitted during the decay of the $^{167}$Yb are indicated. During the decay of the $^{167}$Tm nucleus --- the daughter of the $^{167}$Yb reaction product --- a 207.80\,keV $\gamma$ ray is emitted, figure B shows this transition (together with the 208.95\,keV background transition emitted during the decay of the $^{67}$Ge produced via $^{64}$Zn($\alpha$,n)$^{67}$Ge reaction) and figure C presents its decay curve. The $\gamma$-lines used to determine the cross section of the $^{166}$Er($\alpha$,n)$^{169}$Yb are marked on the lower figure. For further details, see text.}
\end{figure*}
\end{center}

The thickness of the targets varied between 114 and 238\,$\mu$g/cm$^2$, corresponding to an $\alpha$ energy loss of $\Delta$E $\approx$\,15\,keV
(at E$_{\alpha}$\,=\,17\,MeV) and $\Delta$E $\approx$\,39\,keV (at E$_{\alpha}$\,=\,12\,MeV), respectively.
The $\alpha$ energy loss was calculated using the SRIM \cite{srim} code. Thicker targets were used at low bombarding energy, where the cross section and the corresponding $\gamma$ yield is smaller.

\subsection{Irradiation and $\gamma$ counting}
\label{sec:ir}

The erbium targets were irradiated with $\alpha$ beams from the
MGC 20 cyclotron of Atomki. Energies were in the range of
12\,$\leq$\,E$_{\alpha}$\,$\leq$\,17\,MeV, covered in steps of about 0.5\,MeV.
After the beam-defining aperture, the chamber was insulated and a secondary electron suppression voltage of -300\,V was applied at the entrance of the chamber. From the last beam-defining aperture the whole chamber served as a
Faraday cup. The collected charge was measured with a current integrator, the counts were
recorded in multichannel scaling mode, stepping the channel in every minute to take into account the possible changes in the
beam current. The length of the irradiation were between 7 and 25 hours corresponding to number of incident $\alpha$ particles in each irradiation between 0.9\,$\times$\,10$^{17}$ and 6.1\,$\times$\,10$^{17}$.

Before the experiment several beam tests were performed to determine the maximum tolerable $\alpha$ beam current. These tests showed that there was no deterioration of the targets using an $\alpha$ beam current less than 2.5\,$\mu$A. The target thickness was monitored by measuring the yield of the back-scattered $\alpha$ particles, for this purpose an ion implanted silicon detector built into the chamber at $\theta$\,=\,165$^{\circ}$ relative to the beam direction was used. The spectra of this detector were recorded regularly and the ratio of the alpha backscattering yields divided by the corresponding current integrator counts were fitted using a linear function. It was found that the target-losses are typically in the order of 0.6\% and the maximum target loss was 1.2 $\pm$ 0.15\%. Figure \ref{fig:rbs} shows a typical alpha backscattering spectrum (A) and the corresponding stability curve (B) with the fitted linear function.

\begin{table}
\center
\caption{\label{tab:decay}
Decay parameters of the $^{164}$Er($\alpha$,n)$^{167}$Yb and $^{166}$Er($\alpha$,n)$^{169}$Yb reaction products taken from the literature \cite{nds_167Yb, nds_169Yb}.}
\begin{tabular}{cccccc}
\parbox[t]{1.4cm}{\centering{Reaction}} &
\parbox[t]{1.4cm}{\centering{Threshold \\ (MeV)}} &
\parbox[t]{0.8cm}{\centering{Product \\ nucleus}} &
\parbox[t]{1.8cm}{\centering{Half-life \\ (hour)}} &
\parbox[t]{2.4cm}{\centering{$\gamma$-ray \\energy (keV)}} &
\parbox[t]{3.8cm}{\centering{Relative $\gamma$-intensity \\ per decay (\%)}} \\
\hline
$^{164}$Er($\alpha$,n)$^{167}$Yb&11.27&$^{167}$Yb  &  0.292\,$\pm$\,0.003 & 106.2 & 22.44\,$\pm$\,1.35 \\
&&            &                    & 113.3 & 55.08\,$\pm$\,2.97 \\
&&            &                    & 176.2 & 20.40\,$\pm$\,0.80  \\
&&$^{167}$Tm  &  222  \,$\pm$\,4.8  & 207.8 & 41.54\,$\pm$\,8.00 \\
$^{166}$Er($\alpha$,n)$^{169}$Yb&10.44&$^{169}$Yb  &  768.4\,$\pm$\,1.2  & 63.1 & 43.62\,$\pm$\,0.23 \\
&&&&109.8 & 17.39\,$\pm$\,0.09 \\
&&&&130.5 & 11.38\,$\pm$\,0.05 \\
&&&&177.2 & 22.28\,$\pm$\,0.11 \\
&&&&197.9 & 35.93\,$\pm$\,0.12 \\
\hline
\end{tabular}
\end{table}

After the irradiations, 0.25\,h waiting time was used in order to let short-lived activities, which would impact the quality of the measurement, decay. The duration of the $\gamma$ countings were about 200 - 300\,h in case of each irradiation. The decay parameters of the reaction products are given in Table~\ref{tab:decay}.
To determine the $^{164}$Er($\alpha$,n)$^{167}$Yb reaction cross section first the decay of the produced $^{167}$Yb (T$_{1/2}$ = 0.292 $\pm$ 0.003 h) was followed, the yield of the 106.2\,keV and 113.3\,keV $\gamma$ transitions were measured. Moreover, the $^{167}$Tm nucleus (T$_{1/2}$ = 222 $\pm$ 4.8 h), the daughter of the produced unstable $^{167}$Yb, decays by electron capture to $^{167}$Er with emission of 207.8\,keV $\gamma$ ray, which was also used to determine the reaction cross section. The determination of the $^{166}$Er($\alpha$,n)$^{169}$Yb reaction cross section is based on the measurement of the $\gamma$ rays emitted during the $\beta$ decay of the produced $^{169}$Yb (e.g. see figure~\ref{fig:spectra}).

\begin{table*}
\center
\caption{\label{tab:results_164Er}
Measured cross sections of the $^{164}$Er($\alpha$,n)$^{167}$Yb reaction.}
\begin{tabular}{cccc}
\parbox[t]{0.8cm}{\centering{$E_\mathrm{lab}$ [MeV]}} &
\parbox[t]{1.2cm}{\centering{$E_\mathrm{c.m.}$ [MeV]}} &
\parbox[t]{3.0cm}{\centering{Cross section [mbarn]}} \\
\hline
13.5 &13.17     $\pm$   0.04   &       0.069 $\pm$     0.006   \\
14.0 &13.66     $\pm$   0.04   &       0.208   $\pm$   0.018   \\
14.5 &14.14     $\pm$   0.07   &       0.665   $\pm$   0.058   \\
14.51&14.16     $\pm$   0.15   &       0.779   $\pm$   0.079   \\
15.0 &14.63     $\pm$   0.08   &       1.55    $\pm$   0.15    \\
15.5 &15.12     $\pm$   0.05   &       4.62    $\pm$   0.41    \\
16.0 &15.62     $\pm$   0.08   &       10.3    $\pm$   0.92    \\
16.5 &16.09     $\pm$   0.08   &       18.6    $\pm$   1.6     \\
16.52&16.11     $\pm$   0.05   &       20      $\pm$   1.8     \\
17.0 &16.59     $\pm$   0.05   &       29      $\pm$   2.7     \\
\hline
\end{tabular}
\end{table*}

For the $\gamma$ counting a low-energy photon spectrometer (LEPS) was used. In order to reduce the laboratory background a multilayer quasi 4$\pi$ shield was built around
the LEPS detector including an inner 4 mm thick layer of copper, a
2 mm thick layer of cadmium, and an 8\,cm thick outer lead shield \cite{szucs_st}.
To limit the systematic uncertainty the activity of several samples irradiated at 15.0, 16.0 and 17.0\,MeV were measured using a coaxial HPGe detector with 100\,\% relative efficiency, placed in a low background shielding, too. The resulted cross sections are within 5.2\,\% compared to the ones based on the counting with the LEPS detector.

In the case of both detectors, the distance between the target and the end cup was
1\,cm. The detector efficiencies for both detectors used in the
present work had to be known in this close geometry with
high precision. A similar procedure was used to derive
the photopeak efficiencies in the case of the coaxial HPGe and
LEPS detectors. First the absolute detector efficiency was
measured in far geometry: at 15\,cm distance from the
surface of the detector, using calibrated $^{57}$Co, $^{133}$Ba,
$^{137}$Cs, $^{152}$Eu, and $^{241}$Am sources.
Since the calibration sources (especially $^{133}$Ba and $^{152}$Eu)
emit multiple $\gamma$ radiations from cascade transitions, in the
close geometry a strong true coincidence summing effect is
expected, resulting in an increased uncertainty of the measured
efficiency. Therefore, no direct efficiency measurement in
close geometry was carried out. Instead, the activity of several
irradiated erbium targets were measured in both close and far
geometries. Taking into account the time elapsed between
the two countings, a conversion factor of the efficiencies between the two geometries could be determined and used
henceforward in the analysis.

\section{Results and discussion}
\label{sec:res}

\begin{table*}
\center
\caption{\label{tab:results_166Er}
Measured cross sections of the $^{166}$Er($\alpha$,n)$^{169}$Yb reaction.}
\begin{tabular}{cccc}
\parbox[t]{0.8cm}{\centering{$E_\mathrm{lab}$ [MeV]}} &
\parbox[t]{1.2cm}{\centering{$E_\mathrm{c.m.}$ [MeV]}} &
\parbox[t]{3.0cm}{\centering{Cross section [mbarn]}} \\
\hline
12.00   &       11.70   $\pm$   0.07    &       0.0028  $\pm$   0.0004    \\
12.50   &       12.18   $\pm$   0.07    &       0.011   $\pm$   0.001     \\
13.00   &       12.68   $\pm$   0.04    &       0.027   $\pm$   0.003     \\
13.50   &       13.17   $\pm$   0.04    &       0.104   $\pm$   0.009     \\
14.00   &       13.66   $\pm$   0.04    &       0.264   $\pm$   0.023     \\
14.51   &       14.16   $\pm$   0.15    &       0.89    $\pm$   0.08      \\
15.00   &       14.63   $\pm$   0.08    &       2.26    $\pm$   0.19      \\
15.01   &       14.64   $\pm$   0.08    &       2.06    $\pm$   0.17      \\
15.50   &       15.12   $\pm$   0.05    &       5.21    $\pm$   0.44      \\
16.00   &       15.62   $\pm$   0.08    &       11.3    $\pm$   0.96      \\
16.50   &       16.11   $\pm$   0.05    &       21.2    $\pm$   1.79      \\
17.00   &       16.59   $\pm$   0.05    &       37.2    $\pm$   3.17      \\
17.01   &       16.61   $\pm$   0.05    &       35.9    $\pm$   3.17      \\
\hline
\end{tabular}
\end{table*}

The measured ($\alpha$,n) cross section values are listed in
Tables~\ref{tab:results_164Er} and \ref{tab:results_166Er}. The quoted uncertainty in the $E_\mathrm{c.m.}$ values
corresponds to the energy stability of the $\alpha$ beam and to
the uncertainty of the energy loss in the target, which was
calculated using the SRIM code \cite{srim}. The
uncertainty of the cross sections is the quadratic sum of the
following partial errors: efficiency of the detectors (6\,\% (HPGe) and 5\,\% (LEPS), respectively), number of target atoms (4.4\,\%),
current measurement (3\,\%), uncertainty of decay parameters
($\leq$\,8.0\,\%), and counting statistics (0.4\,\% - 7.4\,\%). Some irradiations were
repeated at the same energies. The cross sections were then
derived from the averaged results of the irradiations weighted
by the statistical uncertainty of the measured values.

\begin{center}
\begin{figure*}
\resizebox{1.0\columnwidth}{!}{\rotatebox{0}{\includegraphics[clip=]{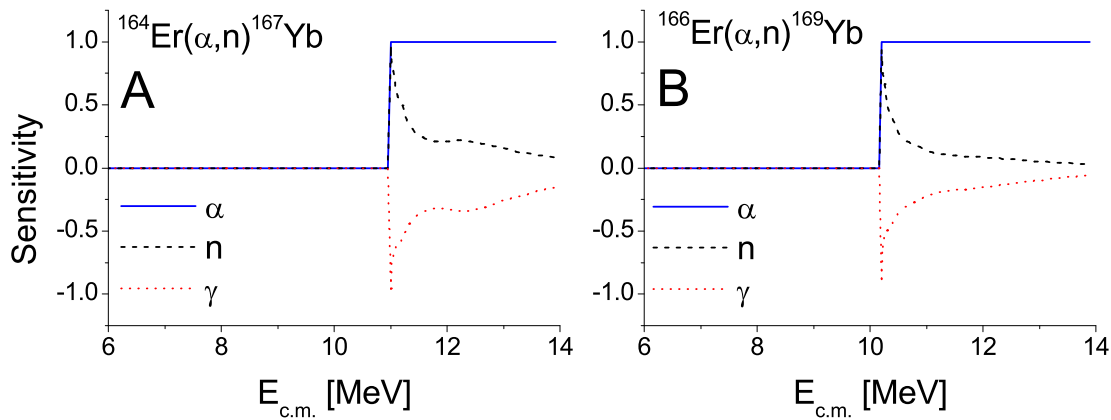}}}
\caption{\label{fig:sensi} (Color online) Sensitivities of the investigated reaction cross sections (A: $^{164}$Er($\alpha$,n)$^{167}$Yb, B: $^{166}$Er($\alpha$,n)$^{169}$Yb) to the $\alpha$, $\gamma$ and neutron widths. For details, see text.  }
\end{figure*}
\end{center}

The prediction of averaged reaction channel widths is crucial for the Hauser-Feshbach model of nuclear reactions \cite{hau52,rau11}. In this well-established model it is assumed that a large number of unresolved resonances is located around the formation energy $E_\mathrm{C}=E_\mathrm{c.m.}+E_\mathrm{sep}^\alpha$ of the compound nucleus (in our case $^{168,170}$Yb), with $E_\mathrm{sep}^\alpha$ being the $\alpha$-separation energy in the compound. Instead of individual resonance widths, an average over all resonance widths is used, leading to averaged neutron-, proton-, $\alpha$-, and $\gamma$-widths (further widths are negligible at the investigated energies). These widths, describing the formation and decay of the compound nucleus states are predicted from further models. Following \cite{rau_sensi,rau14}, the sensitivity 

\begin{equation}
s=\frac{\Gamma}{\sigma}\frac{d\sigma}{d\Gamma} \quad,
\end{equation}

describes the change in the cross section $d\sigma$ when a width $\Gamma$ is changed by $d\Gamma$. Not all the widths are important at each energy. The sensitivity of the calculated cross sections to variations of the different widths is plotted in Figure~\ref{fig:sensi} (the proton width is not shown as the cross section is insensitive to it across all shown energies). A sensitivity of unity means that the cross section is changed by the same factor $f$ as the input parameter, whereas the cross section does not change when the sensitivity is zero. As can be seen the cross sections depend dominantly on the $\alpha$-width at the upper end of the measured energy range. This is because the averaged $\alpha$-width remains much smaller than the neutron width at energies well above the reaction threshold. Towards lower energy approaching the reaction threshold, the neutron width strongly decreases and the cross sections become increasingly sensitive to the neutron- and $\gamma$-widths in addition to the $\alpha$-widths. It was already pointed out in \cite{glo14} that this behavior does not allow for an unambiguous identification of the source of discrepancy closer to the threshold from ($\alpha$,n) data alone, if such a discrepancy between cross section measurements and prediction were found.

\begin{center}
\begin{figure*}
\resizebox{1.0\columnwidth}{!}{\rotatebox{0}{\includegraphics[clip=]{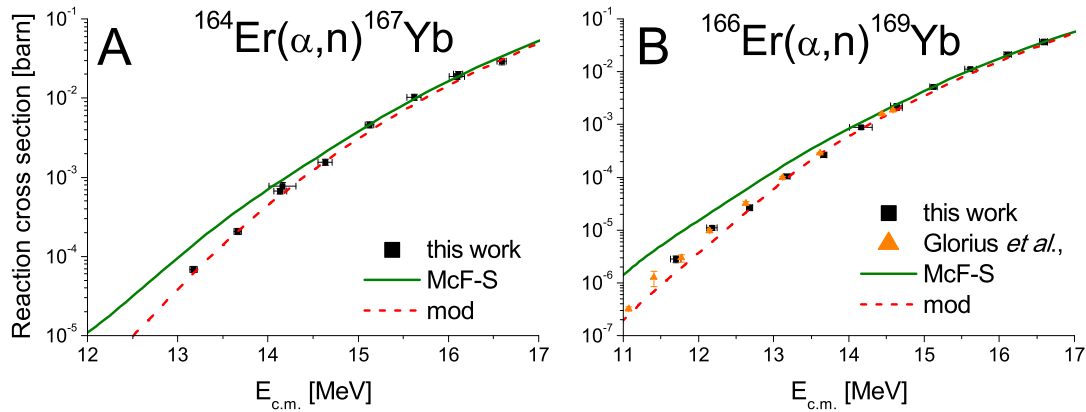}}}
\caption{\label{fig:hkm} (Color online) Measured reaction cross sections of the $^{164}$Er($\alpha$,n)$^{167}$Yb (A) and $^{166}$Er($\alpha$,n)$^{169}$Yb (B) reactions compared to Hauser-Feshbach predictions (see text).}
\end{figure*}
\end{center}

A similar situation as in \cite{glo14} is encountered here. Figure~\ref{fig:hkm} shows a comparison of our data to Hauser-Feshbach predictions performed with the code SMARAGD \cite{smaragd}. In the case of the $^{166}$Er($\alpha$,n)$^{167}$Yb reaction, figure~\ref{fig:hkm} also includes the previous data by \cite{glo14} which are in good agreement with the data obtained in this work. While there is excellent agreement of the data with the standard calculation (marked by the full line and denoted by ``McF-S'' in the figures) at higher energy, an increasing deviation is found towards lower energy for both reactions. This indicates that the $\alpha$-width is described well at the higher energies. The main ingredient for the calculation of the averaged $\alpha$-width is the $\alpha$+nucleus optical potential, for which the global potential by McFadden and Satchler \cite{mcf} was used. The discrepancies towards lower energy cannot be unambiguously assigned to a misprediction of a certain width when solely using the present ($\alpha$,n) data.

A recent measurement provided cross sections for both $^{162}$Er($\alpha$,n)$^{165}$Yb and $^{162}$Er($\alpha$,$\gamma$)$^{165}$Yb reactions, with ($\alpha$,$\gamma$) data also below the ($\alpha$,n) threshold \cite{kis14}. Due to the additional ($\alpha$,$\gamma$) data it was possible to show in \cite{kis14} that an energy-dependent modification of the optical $\alpha$+nucleus potential is required at low energies. Additionally, close to the ($\alpha$,n) threshold the ratio of the neutron- to $\gamma$-width had to be increased by about 40\% (without the possibility to assign this change to either width). For comparison, in Figure~\ref{fig:hkm} we show also cross sections obtained from calculations using the same modifications of widths as in \cite{kis14} (dashed lines labelled ``mod''). Again, good agreement is obtained without further changes. It should be noted that a modification of the 
neutron- to $\gamma$-width ratio only affects the cross sections very close above the reaction thresholds (up to about 0.7 MeV above threshold) because of the strongly decreasing sensitivity to those widths with increasing energy, which can be seen in Figure~\ref{fig:sensi}. Therefore, the changes in the cross sections in the energy ranges covered by the present measurements, and shown in Figure~\ref{fig:hkm}, are solely due to the modification of the $\alpha$ widths.

\section{Summary}
\label{sec:sum}

For the first time, ($\alpha$,n) cross sections of the $p$ nucleus $^{164}$Er were measured at energies well below the Coulomb barrier, using the activation technique. Furthermore, the $^{166}$Er($\alpha$,n)$^{169}$Yb reaction was studied at energies overlapping with the results of a recent measurement performed by \cite{glo14}. The agreement between the two data sets is almost perfect, this fact strengthens further the conclusions of \cite{glo14}.

The experimental data were compared to statistical model predictions. Good agreement was found at the upper end of the measured energy range when using the global $\alpha$+nucleus optical potential by \cite{mcf} but deviations between data and calculations were found towards the reaction threshold in both reactions. Using the same set of parameters as determined in a previous, simultaneous measurement of $^{162}$Er($\alpha$,$\gamma$) and $^{162}$Er($\alpha$,n) \cite{kis14} works well also in the present cases. This confirms the necessity of an energy-dependent modification of the $\alpha$+nucleus optical potential at very low energies.

The present measurements enlarge the database for deriving an improved global $\alpha$+nucleus optical potential at very low energies by future theoretical models. The relevant astrophysical energy range, however, is below the reaction threshold and therefore further information on the energy dependence of the $\alpha$ widths at astrophysical energies requires additional experiments.

\pdfbookmark[1]{Acknowledgments}{sec:ack}
\ack
\label{sec:ack}

This work was supported by OTKA (K101328, PD104664, K108459) and by the ENSAR/THEXO European FP7 programme.
G. G. Kiss acknowledges support from the J\'anos Bolyai Research Scholarship of the Hungarian Academy of Sciences. T. Rauscher acknowledges support from the UK STFC
grant BRIDGCE (ST/M000958/1), the Swiss NSF and the European Research Council.

\section*{References}
\label{sec:ref}
\bibliographystyle{unsrt}



\end{document}